\documentclass[aip,pop,psfig,onecolumn,showpacs,superscriptaddress,reprint,floatfix]{revtex4-1}
\usepackage{amsmath}
\usepackage{mathtools}
\usepackage{amssymb}
\usepackage{color}
\usepackage{graphics}
\usepackage{epsfig}
\usepackage[normalem]{ulem} 
\usepackage{cancel}
\usepackage[mathlines]{lineno}
\graphicspath{{Figures/}}
\usepackage{xtab,afterpage,longtable}
\usepackage{booktabs}   
\usepackage{ltablex}
\relax
\usepackage{float}

\begin{document}
\title{Nonlinear dispersion relation of dust acoustic waves using the Korteweg-de Vries model}
\author{Farida Batool}
\email{farida.batool@iitjammu.ac.in}
\affiliation{Indian Institute of Technology Jammu, Jammu, J\&K, 181221,  India}
\author{Ajaz Mir}
\affiliation{Indian Institute of Technology Jammu, Jammu, J\&K,  181221, India}
\affiliation{Institute for Plasma Research, Gandhinagar, Gujarat, 382428, India}
\author{Sanat Tiwari}
\email{sanat.tiwari@iitjammu.ac.in}
\affiliation{Indian Institute of Technology Jammu, Jammu, J\&K, 181221,  India}
\author{Abhijit Sen}
\affiliation{Institute for Plasma Research, Gandhinagar, Gujarat, 382428, India}
\date{\today}
\begin{abstract}
In this brief communication, we present an exact analytic nonlinear dispersion relation (NLDR) for the dust acoustic waves using the Korteweg-de Vries (KdV) model. The NLDR agrees with the spectrum of spatio-temporal evolution obtained from an exact solution as in Mir~\textit{et al.}~[Phys. Plasmas \textbf{27}, 113701 (2020)]. The NLDR also shows a reasonable match with the experimental data of Thompson~\textit{et al.}~[Phys. Plasmas \textbf{4}, 2331 (1997)] in the long wavelength limit ($k \lambda_D \ll 1$). We suggest that such nonlinear corrections  should be incorporated in the dispersion relation along with damping, streaming, and correlation effects in order to provide a more realistic interpretation of experimental data.
\end{abstract}
\maketitle
\section{Introduction}
\label{intro}
\paragraph*{}
Dust acoustic waves (DAWs) are analogs of ion acoustic waves in a dusty plasma that exist due to the balance of charged dust inertia and plasma pressure~\citep{Rao_PSS_1990, Shukla_IOP_2001}. These waves can arise naturally~\cite{Heinrich_PRL_2009, Merlino_POP_2009, Flanagan_POP_2010, Tadsen_POP_2015, Deka_POP_2017} or can be excited by an external perturbation~\cite{Thompson_POP_1997}. They are low-frequency waves in the range of a few tens of Hz with wavelengths of a few tens of mm. Their characteristics slow timescales are due to the heavy mass of the micron sized dust particles. The experimentally observed space-time evolution of DAWs usually shows nonlinear features through the appearance of harmonics in the frequency and wave number domains~\citep{Sachin_AIP_ad_2023, Bandyopadhyay_PRL_2008, Teng_PRL_2009}. The experimental data is frequently compared with theoretically derived linear dispersion  relations that customarily incorporate various linear contributions arising from collisional/kinetic damping, particle streaming and correlation effects. However since experimental data is based on measurements of finite amplitude waves its nature cannot be fully captured in a linear model based dispersion relation. In this paper we propose a nonlinear dispersion relation that is based on the KdV model and that can provide a better description of the DAW data in the weakly nonlinear dynamical regime.
\paragraph*{}
DAWs show a sound wave nature in the long-wavelength limit~\cite{Merlino_POP_2009, Flanagan_POP_2010}, a dispersive nature for short wavelengths and a phase velocity reversal in the strongly coupled regime~\cite{Pieper_PRL_1996}. As one of the fundamental modes of a dusty plasma, it has been an object of intense research ever since its theoretical prediction by Rao \textit{et al.}~\cite{Rao_PSS_1990} and its experimental identification thereafter by Barkan \textit{et al.}~\cite{Barkan_POP_1995}. Its linear properties were thoroughly explored in weak and strong coupling regimes using fluid models~\cite{kaw_98,diaw_15}, quasi-localized charge approximation~\cite{Rosenberg_PRE_1997, Donko_IOP_2008}, molecular dynamics simulations~\cite{Ohta_PRL_2000, Donko_IOP_2008} and in laboratory experiments~\cite{Barkan_POP_1995}.
\paragraph*{}
A simple linear dispersion for DAW, $\omega/k \approx \lambda_D \omega_{pd} = C_{pd}$, was proposed by Rao \textit{et al.}~\cite{Rao_PSS_1990} in the long wavelength limit $k \lambda_D \ll 1$. Here, $C_{pd}$ is the dust sound speed in the medium, $\lambda_D$ is the Debye length due to Boltzmann ions and electrons, and $\omega_{pd}$ is the characteristic frequency of DAW.
For experimental conditions effects due to dust-neutral collisions, ion streaming, and other damping mechanisms can be significant. Ruhunusiri \textit{et al.}~\cite{Ruhunusiri_POP_2014} have extensively explored the dispersion relation of DAW by retaining most of the above mentioned effects except nonlinear corrections arising from the finite size of the wave amplitude. In subsequent work, \citet{Goree_POP_2020} provided a kinetic dispersion relation for DAWs and compared it with data from the Plasma Kristall-4 (PK-4) experiment on the International Space Station. Interestingly, we found that, this experimental data also shows features arising from non-linearity of the propagating wave. However, such nonlinear effects have not been incorporated so far in the context of a dispersion relation. 
\paragraph*{}
The Korteweg-de Vries (KdV) equation successfully models low frequency weakly nonlinear wave propagation in a variety of media including dusty plasmas \citep{ Debnath_SP_2005,Karczewska_PRE_2014}. The conoidal waves and solitons are exact solutions of the KdV equation and have found applications in domains like hydrodynamics~\citep{Drazin_CUP_1989}, oceanography~\citep{Costa_PRL_2014}, plasmas~\citep{Shukla_IOP_2001}, nonlinear optics~\citep{Leblond_PRA_2008} and astrophysical systems~\citep{Coutant_PRD_2018}. Optical fiber communication is one such practical application where signals in the form of soliton pulses are used to transmit information over long distances without suffering any distortion or dissipation.
While the KdV equation, derived using a reductive perturbation technique, represents a reduced dynamical form of the full fluid model, it retains the essential features of nonlinearity and dispersion that influence the wave propagation. We have utilized the fact that an exact solution is possible for the KdV equation to obtain an exact nonlinear dispersion relation (NLDR) for DAW.
\paragraph*{}
While the linear dispersion relation (LDR) for KdV is trivial, the wave's phase velocity and modes interaction drastically change in the nonlinear regime. A typical perturbation grows in amplitude and interacts nonlinearly with others due to the spatiotemporal term of form $n (x,t)\partial n(x,t)/\partial x$ ( a one-dimensional model). Due to the nonlinear interaction, there is a generation of harmonics of fundamental modes in the medium that can be seen in the power spectrum of the exact analytic solution and the numerical solution of the KdV equation~\citep{Ajaz_POP_2022}. In this case, the functional form of the dispersion relation appears to change from $\omega = f(k)$ to $\omega = F(k,\kappa)$. Here, $\omega$, $k$, and $\kappa$ are the frequency, wave number, and the nonlinearity parameter associated with the cnoidal wave~\citep{Ajaz_POP_2020}. This paper has derived this functional form $F$ for the nonlinear dispersion relation.
\paragraph*{}
Various generalizations of the KdV model exist in the literature. These models are obtained by including the medium's different dissipative or dispersive mechanisms. As an example, the viscous damping leads to the KdV-Burgers equation~\cite{Ajaz_PRE_2023}. Contrary to the KdV equation, there is no known exact solution to these extended model forms of KdV. In all those cases, the numerical evolution data reflects a change in the frequency due to the nonlinearity. Our results in this paper provide a benchmark by comparing the obtained NLDR with those obtained from exact analytic and numerical solutions for the KdV model. The obtained NLDR also facilitates the estimation of the phase velocity of a cnoidal or cnoidal-like nonlinear traveling solution. 
\paragraph*{}
The paper is organized as follows: The KdV equation, its linear dispersion relation, analytic solution, and numerical solution have been discussed in section~\ref{KdV_model}. 
\textcolor{black}{The NLDR from the analytic and numerical spatiotemporal evolution of the KdV equation and the NLDR for the modified KdV (mKdV) equation have been discussed in section~\ref{NLD_KdV}.} 
The theoretical NLDR from the KdV model has been compared to the experimental observation of DAW in the long wavelength limit in section~\ref{DAW_disp}. The work is summarized in section~\ref{sum} with a mention of future scope.
\section{Korteweg-de Vries model}
\label{KdV_model}
\paragraph*{}
The KdV equation represents the weakly nonlinear dynamics of fluids in the absence of dissipating mechanisms and is written as~\citep{Shukla_IOP_2001, Liu_POP_2018, Sen_ASR_2015, Rao_PSS_1990}
\begin{eqnarray}
\frac{\partial n(x,t)}{\partial t}
+\ 
\alpha\  n(x,t)\frac{\partial n(x,t)}{\partial x} 
+\ 
\beta \frac{\partial^3 n(x,t)}{\partial x^3} 
=\ 
0 .
\label{KdV_eqn}
\end{eqnarray}
\paragraph*{}
Here $n(x,t)$ is the perturbed field of the fluid medium and can be either density, velocity, potential \textit{etc.} $\alpha$ and $\beta$ quantify nonlinearity and dispersion of the medium, respectively. These parameters include all characteristic physical quantities of the medium, including the equilibrium temperature, density, and pressure. 
\textcolor{black}{
The density, length, and time are normalized with the equilibrium dust density, $n_{d0}$, Debye length, $\lambda_D$ and dust plasma period, $\omega_{pd}^{-1}$ respectively. It should also be noted that the perturbed density, $n$, can have values less than 1 for the physical scenario. 
}
The linear dispersion relation for the KdV equation is
\begin{equation}
\omega_{L} = - \beta k^3.
\label{KdV_LD_disp}
\end{equation}
This dispersion relation is in a travelling frame moving in positive x-direction that was used to reduce the full set of fluid equations~\citep{Tiwari_NJP_2012} into the Eq.~\eqref{KdV_eqn}. The frame velocity is chosen as the sound speed in the medium. That transformation shall be included for a complete dispersion relation in a rest frame. We have provided the rest frame dispersion relation of the DAW as a case study in section~\ref{DAW_disp}.
\subsection{Analytic solution of KdV}
\label{Exact_solution_KdV}
\paragraph*{}
The balance between non-linearity and dispersion gives rise to the cnoidal wave solution of the KdV equation and is given by~\citep{Drazin_CUP_1989,Ajaz_POP_2020}.
\begin{eqnarray}
&&
n(x, t) 
=\ 
\mu\ \text{cn}^2 \left[
\frac{\sqrt{\mu \alpha}}{2 \sqrt{\beta \kappa(\kappa +2)}} \xi(x,t) ;\ \kappa  
\right] 
\nonumber 
\\
&& 
\xi(x,t) = 
\left( x - \frac{\kappa + \kappa^2 - 1}{\kappa(\kappa + 2)} \alpha \mu t  \right) ,
\label{KdV_analytic_eqn}
\end{eqnarray}
where $\text{cn}$ and $ \mu$ are Jacobi elliptic function and amplitude of the cnoidal wave, respectively. $ \kappa$ is the nonlinearity index. While $0<\kappa<1$ represents the cnoidal (solitary) wave train, the special case of $\kappa = 1$ gives the single soliton solution.
The solution in Eq.~\eqref{KdV_analytic_eqn} can also be represented in a travelling wave-like form as  given by~\citep{Liu_POP_2018, Dingemans_WS_1997} 
\begin{eqnarray}
&& 
n(x, t) = 
\mu\ \text{cn}^2 \left[
2K(\kappa)\left(\frac{x}{\lambda} - ft\right);\ \kappa  
\right] ,
\label{KdV_CN2_eqn}
\end{eqnarray}
where $K(\kappa)$ is the Jacobi complete elliptic integral of the first kind, $\lambda$ and $f$ are the wave's fundamental wavelength and frequency, respectively. Other secondary modes are the harmonics of the fundamental mode.
\subsection{Numerical evolution of KdV}
\label{Sine_KdV}
\begin{figure}[ht!]
\includegraphics[width = \columnwidth]{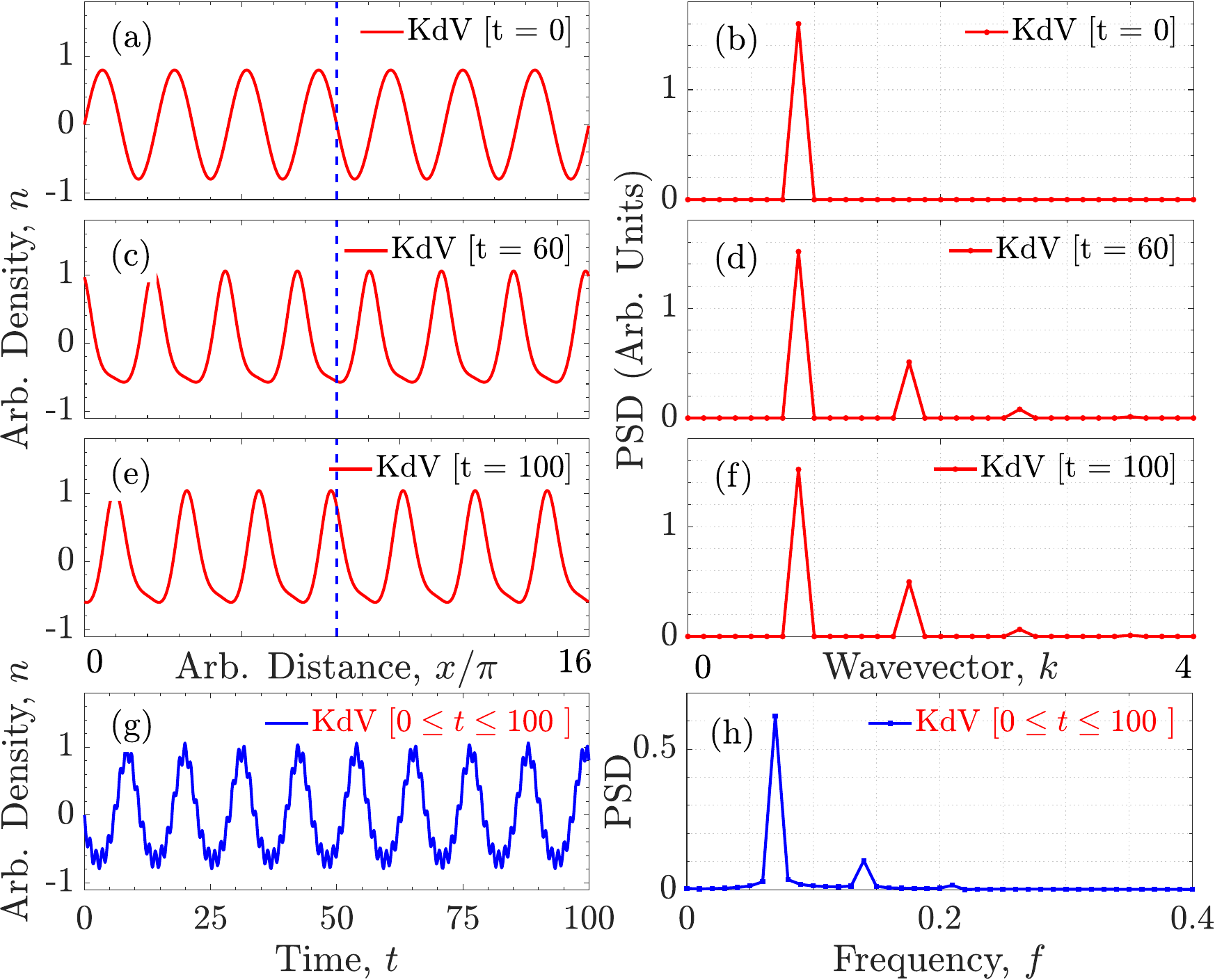}
\caption{Spatio-temporal numerical evolution of sinusoidal perturbation in KdV equation (Eq.~\ref{KdV_eqn}) with $\alpha =2$ and $\beta = 1$. The initial condition is $n(x,0) = A_0\sin(k_0x)$, with $A_0 = 0.8$ and $k_0 = 7k_m$ where $k_m = 2\pi/L_x$ is the minimum wave vector associated with a system of length $L_x = 16\pi$. All parameters are in normalized units.}
\label{Fig_1}
\end{figure}

\paragraph*{}
We have also developed a pseudo-spectral code~\citep{Boyd_DP_2003} using FFTW3 library~\citep{FFTW3_IEEE_2005} to evolve the KdV equation numerically. The code is validated against earlier established results ~\citep{Sen_ASR_2015, Ajaz_POP_2022}. 
Here, we provide a pure sinusoidal perturbation of the form $n(x,0) = A_0\sin(k_0x)$ to the KdV equation. It is then numerically evolved in both space and time. The left column of Figure~\ref{Fig_1}(a, c, and e) shows the spatial profiles of the wave at different times (t = 0, 60, and 100 $\omega_{pd}^{-1}$). The right column of the same figure (b, d, and f) shows the respective power spectral density (PSD), which is obtained by taking the Fourier transform of the corresponding spatial wave profile. We can observe that the initially provided sinusoidal wave distorts into a nonlinear cnoidal waveform at later times. This wave distortion is due to the nonlinear mixing of the modes and, hence, the generation of the harmonics, which we can observe in the PSD. In order to get the temporal profile of the wave, we collected the time series at a particular point in space, represented by subplot (g) and the corresponding Fourier transform  by subplot (h). In the PSD, we observe the generation of harmonics besides the fundamental harmonic.
\paragraph*{}
An analytic solution for the KdV equation is available in Eq.~\eqref{KdV_analytic_eqn}. However, only numerical solutions can be obtained for other generalized KdV models, including additional physical effects. Those numerical solutions can then be Fourier transformed, and a nonlinear dispersion relation can be obtained.
\section{Nonlinear dispersion relation of KdV}
\label{NLD_KdV}
\subsection{Exact nonlinear dispersion relation}
\label{Exact_NLD_KdV}
Comparing the solution of KdV Eq.~\eqref{KdV_analytic_eqn} with Eq.~\eqref{KdV_CN2_eqn}, the wavelength and frequency are given by~\citep{Ajaz_POP_2020}
\begin{eqnarray}
&&
\lambda = 4K(\kappa)\sqrt\frac{\beta(2\kappa + \kappa^{2})}{\alpha \mu}
\label{lambda_eqn_KdV}
\\
&& 
f = \frac{\beta}{4K(\kappa)}\left(\kappa^2+ \kappa -1 \right)\left(\frac{\alpha \mu}{\beta(2\kappa + \kappa^2)}\right)^{3/2}.
\label{freq_eqn_KdV}
\end{eqnarray}
\paragraph*{}
Through mathematical arrangements of Eq.~\eqref{lambda_eqn_KdV} and Eq.~\eqref{freq_eqn_KdV}, the  exact nonlinear dispersion relation for KdV equation can be written as
\begin{equation}
\omega = \frac{4(K(\kappa))^2}{\pi^2}  \left(\kappa^2+ \kappa -1 \right) \beta k^{3} .
\label{KdV_NLD_POP_2022}  
\end{equation}
This relation converts to the linear dispersion relation as in Eq.~\eqref{KdV_LD_disp} in the limit of $\kappa \rightarrow 0$. \textcolor{black}{We have chosen parameters $\alpha = 2.0$, $\beta = 0.0667$, $\kappa = 0.98$, and $\mu = 0.25$ for analytic solution of Eq.~\eqref{KdV_NLD_POP_2022} without the loss of generality. However, choice of these parameter values has been mentioned explicitly as required}.
\begin{figure*}[ht!]
\includegraphics[width = \textwidth]{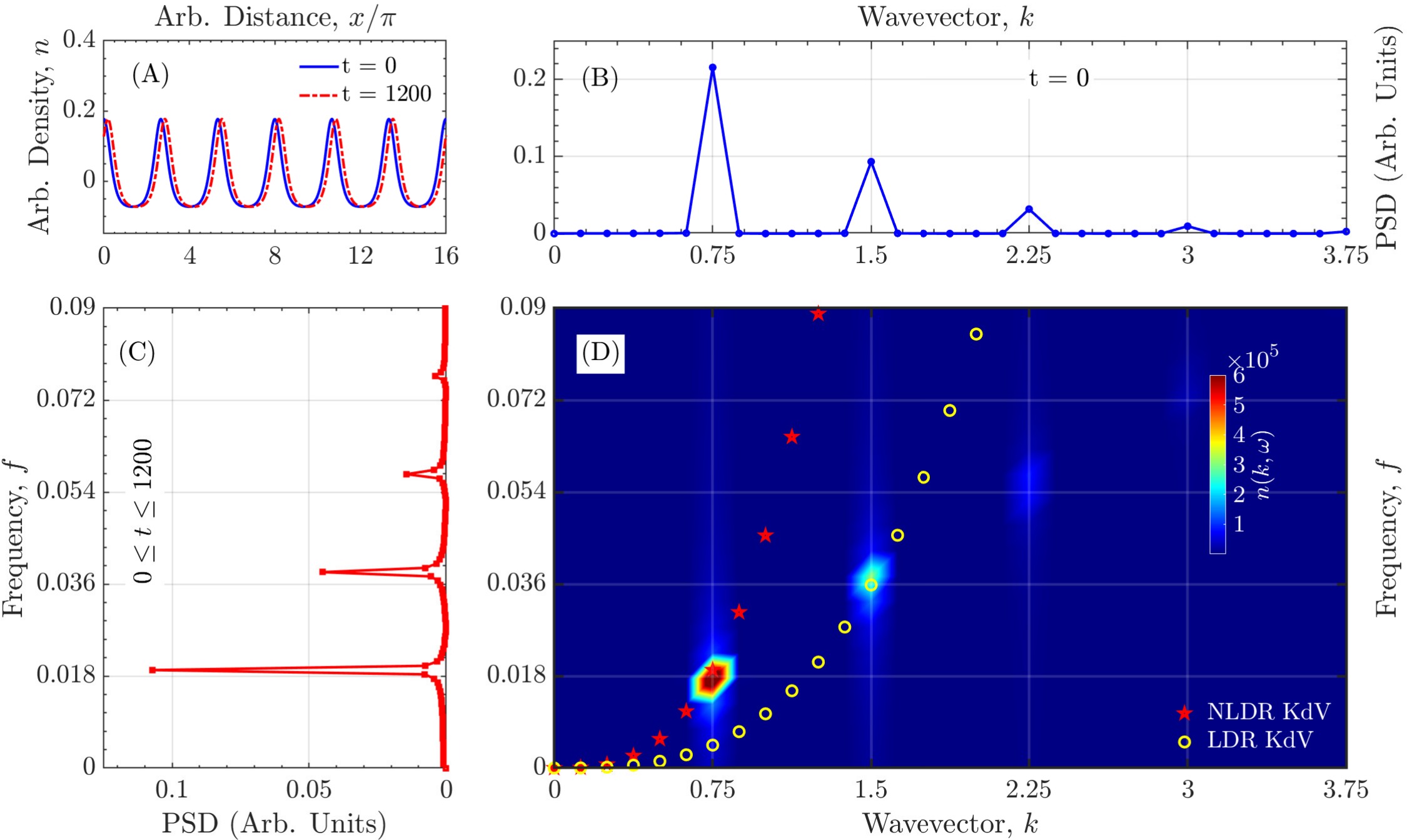}
\caption{The NLDR of KdV (Eq.~\eqref{KdV_eqn}) obtained from analytical solution (Eq.~\eqref{KdV_analytic_eqn}).
(A) The spatial profiles of the analytical solution of KdV at two different times, 0 and $1200$.
(B) and (C) depict PSDs of spatial and temporal profiles, respectively.
(D) represents the comparison of LDR Eq.~\eqref{KdV_LD_disp} (yellow circles), NLDR Eq.~\eqref{KdV_NLD_POP_2022} (red pentagons), and NLDR obtained analytically from the spatio-temporal evolution of KdV (black background with few bright spots).}
\label{Fig_2}
\end{figure*}
\paragraph*{}
We compare the linear and nonlinear dispersion relation to quantify the effect of nonlinearity. For this purpose, we record the spatiotemporal evolution profile from the Eq.~\eqref{KdV_analytic_eqn}. Fig.~\ref{Fig_2} includes all the evolution and dispersion profiles. The subplot (A) shows the spatial evolution of the cnoidal wave at two different times 0 and $1200$ $\omega_{pd}^{-1}$. For a given time of 0 $\omega_{pd}^{-1}$, the PSD of the spatial profile gives the wavenumber associated with the wave and its higher harmonics due to the nonlinearity. \textcolor{black}{This is also true for any subsequent time in evolution including at time of $1200$ $\omega_{pd}^{-1}$. }Further, the subplot (C) shows the fundamental frequency and higher harmonics for PSD of the time series of the cnoidal solution at a given location. Finally, the subplot (D) consolidates the PSD features of subplots (B, C) and compares the linear and nonlinear dispersion relations using Eq.~\eqref{KdV_LD_disp} and Eq.~\eqref{KdV_NLD_POP_2022}, respectively. The NLDR passes through the fundamental frequency, wavenumber (first bright spot in the black background of D) as obtained from Eq.~\eqref{KdV_analytic_eqn}. The extra-bright spots represent the harmonics, which represent the signature of the nonlinearity. Besides this, we can also observe that the NLDR shows an upward shift in the frequency from the LDR.
\subsection{Nonlinear dispersion relation from numerically evolved modified KdV}
\label{nldr_KdV_num}
\begin{figure*}[ht!]
\includegraphics[width = \textwidth]{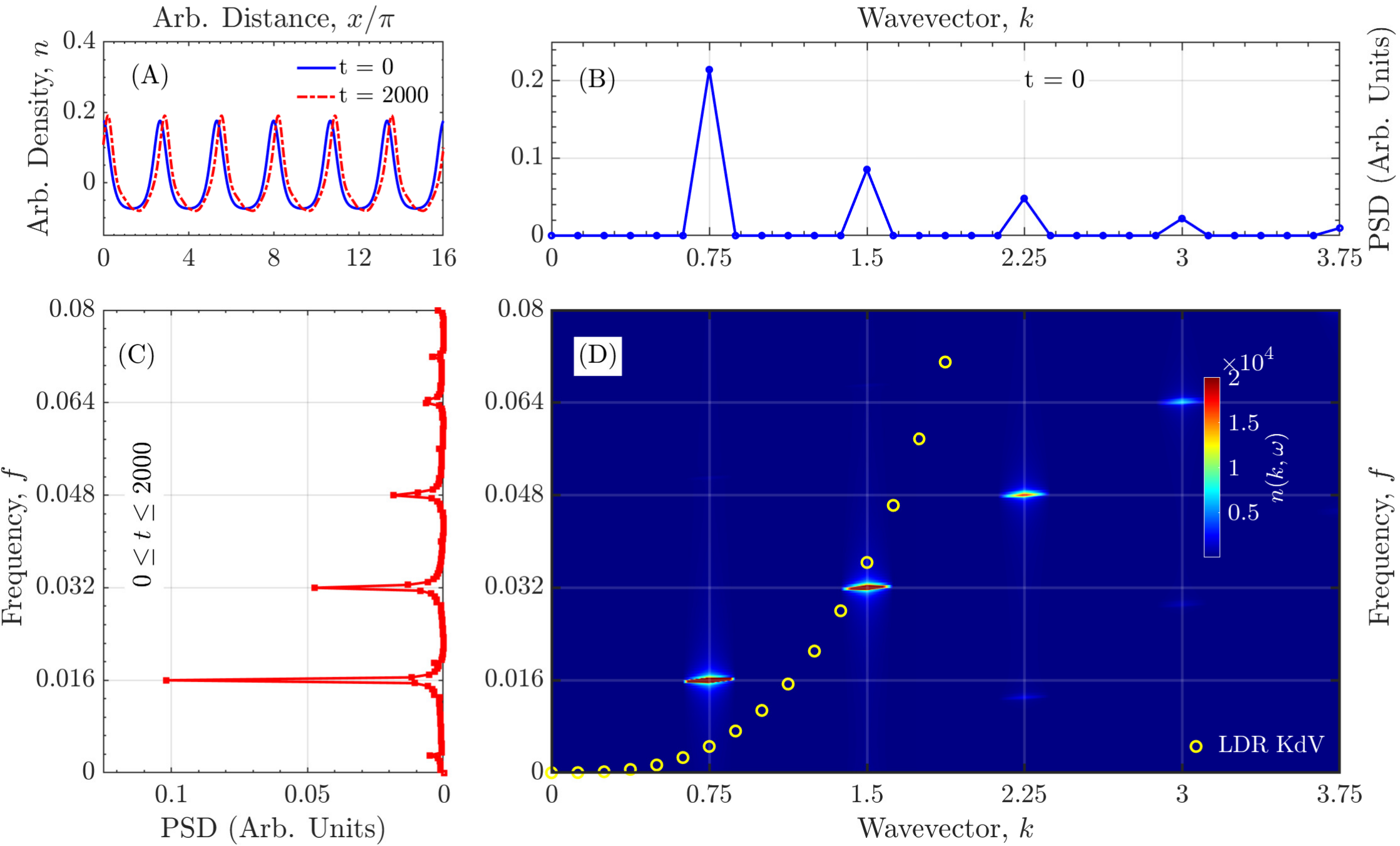}
\caption{
\textcolor{black}{The NLDR of mKdV with nonlinearity of the form $n^2 \partial_x n$. (A) The spatial profiles of the numerical solution of mKdV at two different times, 0 and 2000 with initial perturbation $n(x,0) = A_0 cn^2[2 K(\kappa) x/\lambda; \kappa]$. (B) and (C) depict PSDs of spatial and temporal profiles, respectively. (D) represents the comparison of LDR of Eq.~\eqref{KdV_LD_disp} (yellow circles) and NLDR obtained numerically from the spatio-temporal evolution of mKdV (black background with few bright spots).}
}
\label{Fig_3}
\end{figure*}

\textcolor{black}{
The previous section delineates the importance of considering the effect of nonlinearity while studying the DAW dispersion using the KdV equation and its analytic solution. The effect of nonlinearity on dispersion can be calculated numerically for either different variants of the KdV equation or using the full fluid model. These extended models may incorporate different physical processes such as supra-thermal electrons, relativistic effects, strong coupling, visco-elastic effects, etc.~\citep{Kalita_JPP_2017, Lin_CSF_2007}. These models may not have analytic solutions and hence require solving them numerically. Here, we present one such work by incorporating the dispersion relation obtained by numerically evolving the modified KdV equation. The equation we have chosen differs from the KdV equation (\ref{KdV_eqn}) with the nonlinearity of the form $n^2 \partial n / \partial x = n^2 \partial_x n$. The present dynamical model for dusty plasma can be obtained under the condition of dusty plasmas with variable temperatures~\citep{Kalita_JPP_2017} and due to the non-thermal ions~\cite{Lin_CSF_2007}.
}
\paragraph*{}
\textcolor{black}{
Figure~\ref{Fig_3}~(D) shows the NLDR of mKdV, which is obtained from the numerical spatiotemporal evolution with initial perturbation of the from $n (x,0) = A_0 cn^2[2 K(\kappa) x/\lambda; \kappa]$. The parameters used for the evolution of the mKdV equation are provided in Table~\ref{Tab_1}. We observe the fundamental frequency for the nonlinear wave to be $0.016$ compared to $0.0045$ for the linear wave corresponding to wavenumber $k$. For reference and code validation purposes, we also evolved the KdV equation numerically using the same initial condition, and their parameters are also mentioned in the same table. From Figure~\ref{Fig_3}~(D), it is evident that there is a frequency increase in comparison to the frequency obtained from the linear dispersion relation. This frequency increase is similar to what we obtained from the KdV model. This, in turn, supports our result that nonlinearity, or all forms due to their physical origins, will show a frequency shift. The magnitude of the frequency shift is governed by the nature of the nonlinearity of the medium for given physical conditions. 
}
\begin{table}[!ht]
\textcolor{black}{
\caption{Parameters for numerical simulations of the KdV $(n \partial_x n)$ and the mKdV $(n^2 \partial_x n)$.}
{\renewcommand{\arraystretch}{1.5}
\begin{tabular}{|p{1.15cm}|p{0.55cm}|p{0.75cm}|p{0.70cm}|p{0.75cm}|p{0.70cm}|p{1.0cm}|p{0.55cm}|p{0.75cm}|}
\hline  
\textbf{Model} & $\alpha$  & $\beta$    & $A_0$  & $\kappa$ & $k$ & $k_m$  & $L_x$  & $f$
\\ \hline 
\textbf{KdV}   & $2$    & $0.067$   & $0.25$    & $0.98$   & $6k_m$  & $2\pi/L_x$  & $16\pi$  & $0.020$
\\ \hline 
\textbf{mKdV}  & $11$    & $0.067$   & $0.25$    & $0.98$   & $6k_m$  & $2\pi/L_x$  & $16\pi$ & $0.016$
\\ \hline 
\end{tabular}
}
\label{Tab_1}
}
\end{table}
\section{Dispersion relation of Dust Acoustic Wave}
\label{DAW_disp}
\paragraph*{}
As a case study, we chose to discuss the nonlinear dispersion relation of DAW. Its linear dispersion relation (within  cold dust approximation) is given by~\citep{Yaroshenko_PRE_2020, Shukla_IOP_2001} 
\begin{equation}
    \omega = kC_{pd}\left[ 1 + (k\lambda_D)^2 \right]^{-1/2} \approx kC_{pd}  - \frac{1}{2} (k\lambda_D)^3 \omega_{pd} .
    \label{DAW_LD_disp}
\end{equation}
Yaroshenko~{\it et al.}~\citep{Yaroshenko_PRE_2020} provided a nonlinear analytic version of the DAW dispersion relation (for cold dust and no charge fluctuations) as:
\begin{eqnarray}
&&
\omega = kC_{pd}\left\{ \left[ 1 + (k\lambda_D)^2 \right]^{-1/2} + \alpha_i  \sqrt{\frac{e\phi}{2 \pi T_i}} \right\}
\nonumber
\\
&&
\omega \approx kC_{pd}\left\{  \left[ 1 - \frac{1}{2} (k\lambda_D)^2 \right] + \alpha_i \sqrt{\frac{e\phi}{2 \pi T_i}}  \right\}
\label{Yaroshenko_NLD}
\end{eqnarray}
Here $\phi$ is the wave amplitude, $e$ is the electron charge, $T_i$ is the ion temperature, and $\alpha_i$ is a parameter that depends on the Jacobi complete elliptic integral of the first $K(\kappa)$ and second $E(\kappa)$ kinds.
In comparison, the nonlinear dispersion relation we obtained from KdV can be written in the stationary frame as:
\begin{equation}
\omega = kC_{pd} \left[1 + \frac{4(K(\kappa))^2}{\pi^2}  \left(\kappa^2+ \kappa -1 \right) \beta \left(k \lambda_D \right)^2 \right]
\label{DAW_NLD_KdV}
\end{equation}
The NLDR of KdV (Eq.~\eqref{DAW_NLD_KdV}) depends only on the Jacobi complete elliptic integral of first kind $K(\kappa)$ and parameter $\beta$. Parameters $\alpha$ and $\beta$ can be calculated from physical plasma parameters as given in Liu~{\it et al.}~\citep{Liu_POP_2018}.
\begin{figure*}[ht!]
\includegraphics[width = \textwidth]{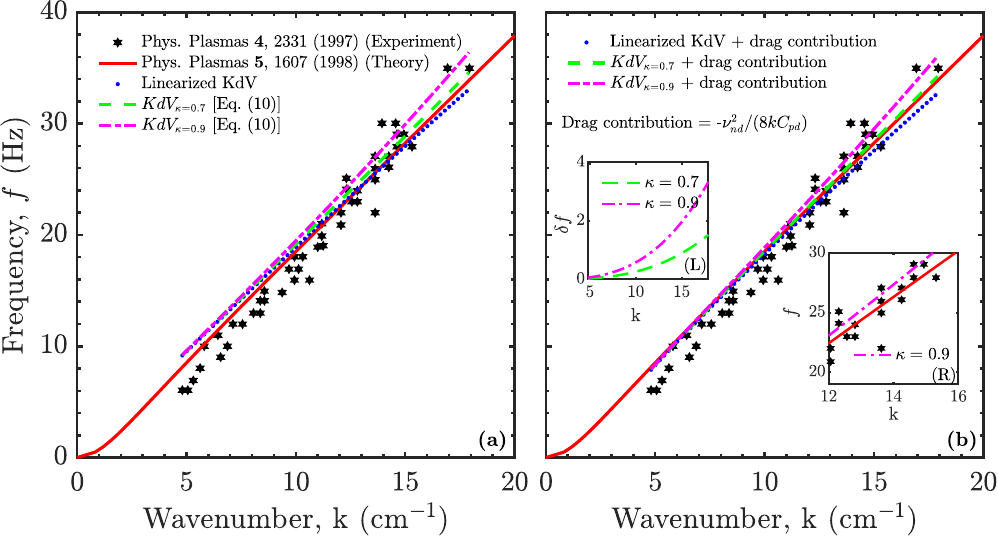}
\caption{Comparison of dispersion relation of DAW, KdV and effect of dust-neutral drag on KdV dispersion relation. 
(a) Experimental DAW (black hexagons)~\citep{Thompson_POP_1997}, theoretical including dust-neutral drag (red bold)~\citep{Merlino_POP_1998}, linearized KdV (black dash-dotted) $\kappa = 0$, NLDR KdV at $\kappa = 0.7$ (green dash-dash) and $\kappa = 0.9$ (magenta dash-dotted) (Eq.~\eqref{DAW_NLD_KdV}).
(b) Effect of drag ($\nu_{nd} = 61\ s^{-1}$) on KdV dispersion relation corresponding to each case as in (a). Inset (L) shows the positive frequency shift as the nonlinearity increased from $\kappa=0.7$ to $0.9$. Inset (R) shows the magnified version of the NLDR of KdV at $\kappa = 0.9$ and the experimental dispersion data. 
}
\label{Fig_4}
\end{figure*}
\subsection{Comparison with experimental observation of DAW in long wavelength limit $k\lambda_{D}<<1$}
\label{comparison_KdV_Exp_disp}
\paragraph*{}
We compare the dispersion relation obtained from Eq.~\eqref{DAW_NLD_KdV} with the experimental result by ~\citet{Thompson_POP_1997} in long wavelength limit $k\lambda_{D}<<1$ and also with the theory by ~\citet{Merlino_POP_1998}, given below. 
\begin{eqnarray}
&&
k= \omega(\omega + \sqrt{(\omega^{2} + \nu_{nd}^{2})})^{1/2}
\label{Theo_disp_1998_daw}
\end{eqnarray}
if $\nu_{nd} /\omega << 1$, the above dispersion relation simplifies as
\begin{eqnarray}
&&
\omega = kC_{pd}(1-\nu_{nd}^2/8k^2C_{pd}^2).
\label{theo_disp_1998_app}
\end{eqnarray}
To get the KdV dispersion (linear and nonlinear) from Eq.~\eqref{DAW_NLD_KdV}, we calculate the value of $\beta$ from experimental parameters using relations derived by~\citet{Tiwari_NJP_2012}. To make a comparison with experimental data, we have added an approximate dust-neutral drag $\nu_{nd}$ as in Eq.~\eqref{theo_disp_1998_app} to the Eq.~\eqref{DAW_NLD_KdV} getting a form:
\begin{equation}
\small{
\omega = kC_{pd} \left[1 + \frac{4(K(\kappa))^2}{\pi^2}  \left(\kappa^2+ \kappa -1 \right) \beta \left(k \lambda_D \right)^2  - \nu_{nd}^2/8k^2C_{pd}^2\right] 
\label{DAW_NLD_KdV_drag}}
\end{equation}
Using the plasma and damping parameters from the reference experimental work, we calculate the NLDR for three values of $\kappa = 0, 0.7$, and $0.9$.
\paragraph*{}
The comparison of experimental data with the proposed theory on NLDR is presented in Fig.~\ref{Fig_4}. The subplot (a) shows the comparison of KdV dispersion with the experiment (black hexagons) and with theory (red bold line, Eq.~\eqref{Theo_disp_1998_daw}). The KdV dispersion relation obtained from Eq.~\eqref{DAW_NLD_KdV} is for three values of nonlinearity parameter $\kappa$ (0, 0.7, and 0.9), which are represented by the black dash-dotted, green dash-dash, and magenta dash-dotted lines, respectively. Please note 
that the dispersion relation represented by Eq.~\eqref{DAW_NLD_KdV} does not include the dust-neutral collision term. To make it further realistic to the reference experimental data, we include dust-neutral collision term from Eq.~\eqref{theo_disp_1998_app} to the  Eq.~\eqref{DAW_NLD_KdV}. The NLDR then takes the form of Eq.~\eqref{DAW_NLD_KdV_drag}. These modified dispersion relations are shown in subplot (b). Adding the drag term, the nonlinear dispersion relation obtained from KdV matches well with the experiment.
\paragraph*{}
The subplot Fig.~\ref{Fig_4}(b) also has left and right insets. The left inset shows the frequency shift for a given wavenumber as the nonlinearity parameter $\kappa$ changes from $0.7$ to $0.9$. The right inset magnifies the dispersion relation, showing a reasonable match of the experimental data with that of the theoretical model at $\kappa = 0.9$.
\section{Summary}
\label{sum}
\paragraph*{}
In this paper, we focus on the role of medium's nonlinearity in estimating the dispersion relation of the waves. We have demonstrated this using the KdV model. Through our analysis, we could show a small positive frequency shift for a wave as it gets nonlinear. At higher wavenumbers, the shift is prominent.
\paragraph*{}
The present work has the convenience of an exact analytic solution for the KdV equation. However, numerical solutions will be necessary for other systems where the dynamics are governed by a modified KdV model or through the full set of fluid equations or lattice base dynamical models.
\textcolor{black}{In this present work, we have shown a positive frequency shift for one such form of mKdV equation by evolving it numerically.}
\paragraph*{}
Finally, connecting our work with dusty plasmas, we derived the nonlinear dispersion relation for DAWs using KdV model and then by transforming it back into the rest frame. We found results qualitatively similar to the dispersion relation proposed by Yaroshenko {\it et al.}~\citep{Yaroshenko_PRE_2020}. We also found a good agreement between our theoretical NLDR and the experimental observations from Thompson \textit{et al.} ~\cite{Thompson_POP_1997} in the long wavelength limit where the KdV-based dispersion relation makes sense. However, we had to add the effect of drag as an artificial term to the NLDR obtained from analytic approaches.
both with and without dust-neutral drag contribution and found a good agreement with their experimental findings.
\begin{acknowledgements}
This work was supported by the Indian Institute of Technology Jammu Seed Grant No. SG0012.
F. B. thanks the University Grants Commission (UGC) of India for the PhD fellowship.
A. S. thanks the Indian National Science Academy (INSA) for the INSA Honorary Scientist position.
We also thank R. Wani for the initial discussions and S. Sharma for providing experimental references on DAWs.
\end{acknowledgements}
\section*{Appendix}
\begin{figure}[hbt!]
\includegraphics [width = \columnwidth]{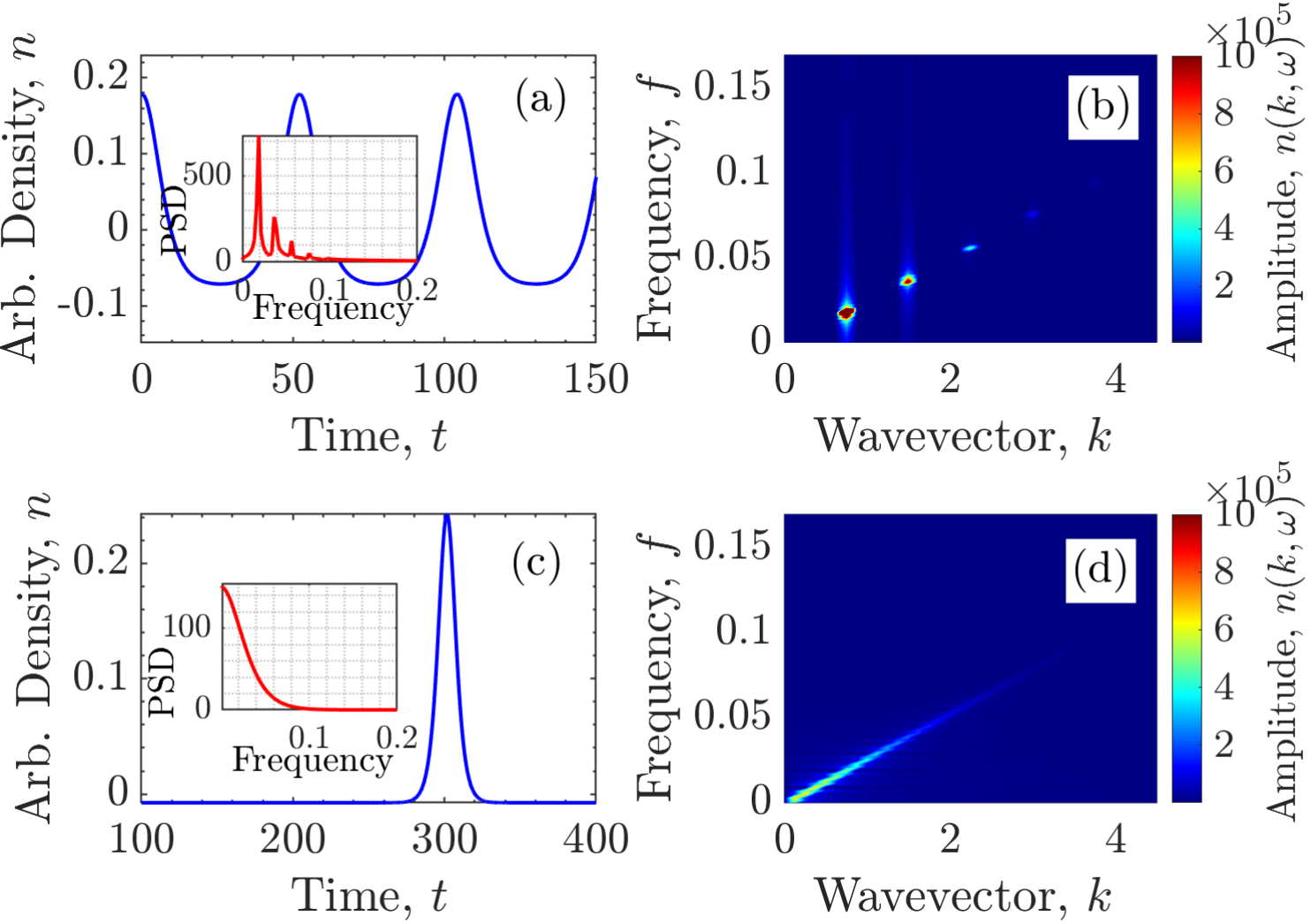}
\caption{Spatio-temporal wavevector-frequency of cnoidal and soliton of KdV Eq.~\eqref{KdV_eqn}.
The time series of (a) cnoidal wave and (c) soliton are the exact solutions of the KdV equation. Insets show the frequency spectrum associated with their time series. The  dispersion relation for cnoidal wave and the soliton are shown in (b) and (d), respectively. The parameters are same as for the Fig.~\ref{Fig_2}.}
\label{Fig_5}
\end{figure}
We provide a comparative picture of the cnoidal and soliton solutions in the time and frequency domains. It is common to see soliton (a single localized structure, Fig.~\ref{Fig_5}(c)) as a limit $\kappa \rightarrow 1$ of the generalized cnoidal wave (a wave-train of multiple localized structures, Fig.~\ref{Fig_5}(a)) with $0 < \kappa <1$. Though it is less intuitive to see how a discrete frequency spectrum (power lies at harmonics, Fig.~\ref{Fig_5}(b)) of a cnoidal wave approaches a continuous spectrum (power lies at all available frequencies, Fig.~\ref{Fig_5}(d)) for the soliton.
\bibliography{NLD_KdV}
\end{document}